\documentclass[12pt]{article}
\usepackage{epsfig}
\usepackage{axodraw}
\def\bea{\begin{eqnarray}}
\def\eea{\end{eqnarray}}
\def\bq{\begin{quote}}
\def\eq{\end{quote}}

\def\gappeq{\mathrel{\rlap
{\raise.5ex\hbox{$>$}}
{\lower.5ex\hbox{$\sim$}}}}
\def\lappeq{\mathrel{\rlap{\raise.5ex\hbox{$<$}}
{\lower.5ex\hbox{$\sim$}}}}
\def\simlt{\stackrel{<}{{}_\sim}}
\def\simgt{\stackrel{>}{{}_\sim}}

\newcommand{\beq}{\begin{equation}}
\newcommand{\eeq}{\end{equation}}
\newcommand{\OO}[1]{\mathbf{\Omega}_{#1}} 
\newcommand{\OD}[1]{\mathbf{\Omega}_{#1}^2} 
\newcommand{\YY}[1]{\mathbf{\tilde{Y}}_\nu^{#1}} 
\def\tmi{\tilde{m}_1}
\def\tim{\tilde{m}_1}

\begin{document}
\pagestyle{empty}
\begin{flushright}
{\bf IFT-03/14\\
\bf hep-ph/0306059\\
\bf \today}
\end{flushright}
\vspace*{5mm}
\begin{center}
{\bf Limits on $T_{\rm reh}$ for thermal leptogenesis with 
hierarchical neutrino masses}
\\
\vspace*{1cm} 
Piotr H. Chankowski and Krzysztof Turzy\'nski 
\\
Institute of Theoretical Physics, Warsaw University, \\
Ho\.za 69, 00-681 Warsaw, Poland 
\vskip 1.0cm

\vspace*{1.7cm} 
{\bf Abstract} 
\end{center}
\vspace*{5mm}
\noindent
{\small We make a simple observation that if one of the right-chiral 
neutrinos is very heavy or its Yukawa couplings to the standard lepton 
doublets are negligible, so that it effectively decouples from the 
see-saw mechanism, the prediction for the baryon asymmetry of the 
Universe resulting from leptogenesis depends, apart from the masses 
$M_1$ and $M_2$ of the remaining two right-chiral neutrinos, only on the 
element $\YY{22}$ of the neutrino Yukawa coupling. For $M_2\simgt10M_1$ 
the lower bound on $M_1$ and also on $T_{\rm reh}$, resulting from the 
requirement of 'successful leptogenesis' is then significantly increased 
compared to the one computed recently by Buchm\"uller {\it et al.} in 
the most general case. Within the framework of thermal leptogenesis, the 
only way to lower this limit is then to allow for sufficiently small mass 
difference $M_2-M_1$.
}
\vspace*{1.0cm}
\date{\today} 


\vspace*{0.2cm}

\vfill\eject
\newpage

\setcounter{page}{1}
\pagestyle{plain}

Accumulated over the past years data on solar and atmospheric neutrino 
oscillations are consistent with the minimal see-saw mechanism of 
neutrino mass generation. Barring the LSND anomaly, all the remaining 
data can be explained by the mixing of three light active neutrino 
species with differences of the masses squared equal to:
\beq
\label{diffsq}
\Delta m_{\rm sol}^2 \approx 7\times 10^{-5}~{\rm eV}^2
\qquad
\Delta m_{\rm atm}^2 \approx 2.5\times 10^{-3}~{\rm eV}^2
\eeq 
and the so-called bi-large pattern of the unitary mixing matrix 
$\mathbf{U}$. In principle, the differences (\ref{diffsq}) 
can result from different patterns of neutrino masses $m_{\nu_i}$ 
satisfying the WMAP bound\footnote{This bound relies on some 
theoretical assumption and can be questioned.} 
$\sum_im_{\nu_i}<0.7$ eV \cite{WMAP}, but the most natural in the context 
of the see-saw mechanism \cite{GERASL} operating at some scale $M$ close 
to the GUT scale $\sim2\times10^{16}$ GeV seems to be the hierarchical 
pattern $m_{\nu_3}\simgt\sqrt{\Delta m_\mathrm{atm}^2}$, 
$m_{\nu_2}\simgt\sqrt{\Delta m_\mathrm{sol}^2}$, 
for which the mixing pattern is stable with respect to radiative corrections 
\cite{CHPO}. In this mechanism, whose main virtue is that it naturally 
explains why neutrinos are so light, three heavy Majorana neutrinos 
$\nu^c_A$ ($A=1,2,3$) couple to the standard $SU(2)$ lepton doublets 
through the Lagrangian term 
${\cal L}=\epsilon_{ij}H_i\nu^c_A\mathbf{Y}_\nu^{AB}l_{jB}+$H.c. 
Below the mass scale of the lightest right-chiral neutrino the 
effective light neutrino mass matrix takes the form
\beq
\label{eqn:seesaw}
\mathbf{m}_\nu=-{v^2\over2}\mathbf{Y}_\nu{\rm ^T}\cdot\mathbf{M}^{-1} 
\cdot\mathbf{Y}_\nu = {v^2\over2M_1}\mathbf{U}^\ast_\nu\cdot 
{\rm diag}(\xi_1^2,\xi_2^2,\xi_3^2)\cdot\mathbf{U}^\dagger_\nu~,
\eeq
where $\mathbf{M}=M_1{\rm diag}(1,x_2^{-2},x_3^{-2})$ is the right-chiral 
Majorana mass matrix, $v=\sqrt2\langle H^0\rangle$ and 
$\mathbf{U}_\nu$ diagonalizes $\mathbf{m}_\nu$.

Another attractive feature of the see-saw mechanism is the possibility 
of explaining the baryon number $B$ asymmetry of the Universe by the out 
of equilibrium lepton number $L$ violating decays of the right-chiral 
neutrinos \cite{FUYA}. The net lepton number generated by these decays 
at temperatures $T\simlt M_1$ is subsequently converted into the baryon 
number by the sphaleron mediated $B+L$ violating transitions. Different
theoretical models of the see-saw mechanism that aim at reproducing 
the experimentally measured neutrino properties in a natural way can 
be therefore further constrained by the requirement of 'successful 
leptogenesis'.

There are two classes of theoretical models of the see-saw mechanism which 
can give bi-large mixing for hierarchical light neutrino mass spectrum 
$m_{\nu_3}\simgt\sqrt{\Delta m^2_{\rm atm}}$, 
$m_{\nu_2}\simgt\sqrt{\Delta m^2_{\rm sol}}$, without too much artificial 
fine tuning of the parameters \cite{ALFE}. These are the so-called 'lopsided' 
models in which large neutrino mixing arises from the charged lepton sector 
\cite{ALBA} and models in which it is due to some special textures of the 
neutrino Yukawa matrix $\mathbf{Y}_\nu$ in the mass eigenstate basis of the 
charged lepton \cite{KI1,KI2,LAMASA,KIRO}. In both types of models the 
see-saw mechanism is usually dominated by contributions of one or at most 
two right-chiral neutrinos, what most naturally happens if their masses 
exhibit some hierarchy $1=x_1>x_2\gg x_3$. One considers also minimal 
see-saw models with exactly two right-chiral neutrinos \cite{RAST}.

The purpose of this note is to investigate leptogenesis in a general 
class of theoretical see-saw models, in which the contribution to the 
see-saw formula (\ref{eqn:seesaw}) of the third right-chiral neutrino is 
negligible. We will be working in the basis, in which the charged lepton 
Yukawa couplings are diagonal and 
therefore $\mathbf{U}=\mathbf{U}_\nu$. As it is well-known \cite{CAIB}, 
the neutrino Yukawa coupling matrix $\mathbf{Y}_\nu$, which enters the 
formulae for the CP asymmetry parameters $\epsilon_i$ \cite{COROVI} and 
various reaction cross sections, cannot be uniquely reconstructed from the 
low energy quantities $m_{\nu_i}$ and $\mathbf{U}_\nu$ and the Majorana 
masses $M_A$. Rather, one has \cite{CAIB,DAIB,BUDIPL,BRetAL}
\beq
\label{eqn:Ygensol}
\YY{AB}\equiv\left(\mathbf{Y}_\nu\mathbf{U}_\nu\right)^{AB} 
=ix_A^{-1}\OO{AB}\xi_B
\eeq
(no summation over $A$ and $B$), where the complex orthogonal matrix 
$\mathbf{\Omega}$ accounts for the six-parameter ambiguity in translating 
the neutrino masses into the neutrino Yukawa couplings 
$\tilde\mathbf{Y}_\nu$. We first point out 
that if one of the right-chiral neutrinos effectively decouples (either 
because it is very heavy or its couplings $\YY{3A}$ are negligible) the 
prediction for the baryon asymmetry of the Universe generated via 
leptogenesis depends only on $M_1$, $x_2$ and $\YY{22}$, that is, the 
six-parameter ambiguity encoded in $\mathbf{\Omega}$ reduces to the 
dependence on two parameters only, which can be chosen as the modulus and 
the phase $\varphi$ of $\YY{22}$. For $M_2\gg M_1$ ($x_2\ll 1$), when the 
lepton number generated by the out of equilibrium decays of the second heavy 
neutrino is completely washed out by the time when the lightest right-chiral 
neutrinos decay, or simply the reheat temperature $T_{\rm reh}$ is too low 
for the second heavy neutrino to be produced thermally, the leptogenesis 
depends only on $M_1$ and the product $x_2\YY{22}$ which can be translated 
into $\varphi$ and the parameter $\tmi$ defined in \cite{PLUMIS}. Compared 
to the analysis of such a scenario performed in \cite{BUDIPL} for the thermal 
production of heavy neutrinos, the assumption about the decoupling of 
one right-chiral neutrino has the following consequences. Firstly, for each 
point of the $(M_1,\tmi,\varphi)$ parameter space the relevant CP asymmetry 
can be unambiguously computed; its maximal value is smaller by about 20\% 
compared to the value $\epsilon_1^\mathrm{max}$ found in \cite{DAIB} and 
adopted in \cite{BUDIPL}. Secondly, and more importantly, the range of 
possible $\tmi$ values is substantially reduced: $\tmi$ is always greater 
than $m_{\nu_2}$ (and not $m_{\nu_1}$ as in \cite{BUDIPL}). This 
significantly increases the lower limit on $M_1$ and, consequently, on the 
reheat temperature $T_{\rm reh}$ for which the leptogenesis mechanism 
dominated by the decays of the lightest right-chiral neutrino can reproduce 
the observed baryon asymmetry of the Universe 
$Y_B\approx(0.8-1)\times 10^{-10}$ \cite{WMAP}. In this letter, for technical 
simplicity we integrate the Boltzmann equations on the non-supersymmetric 
scenario \cite{PLUMIS}, but we believe that our results qualitatively hold 
in the supersymmetric case as well. The lower limit 
$T_{\rm reh}\simgt2\times 10^{12}$ GeV (assuming $T_{\rm reh}\simgt10M_1$), 
which we find in the strict decoupling limit, may then lead to too abundant 
gravitino production, incompatible with the standard nucleosynthesis. The 
only way to lower $T_{\rm reh}$ is to consider almost degenerate two lighter 
right-chiral neutrinos as advocated in \cite{ELRAYA1}. The simplification 
occurring in the decoupling limit allows us to extend the analysis also to 
this case. In particular, we investigate how fast the necessary reheat 
temperature $T_{\rm reh}$ decreases with increasing degeneracy of the two 
lighter right-chiral neutrinos. At the end, we briefly discuss the conditions 
under which the results described here should apply.
\vskip0.3cm

For fixed $\xi_A$ (i.e. for fixed see-saw masses of the left-chiral 
neutrinos), $M_1$, $x_2$ and $x_3$ the formula (\ref{eqn:seesaw}) 
constitutes a set of $2(3+6/2)=12$ real equations allowing to solve 
for six complex elements of $\YY{AB}$ in terms of the remaining three. 
This six-parameter ambiguity is encoded in the complex orthogonal 
matrix $\mathbf{\Omega}$ in eq. (\ref{eqn:Ygensol}). If the third 
right-chiral neutrino is decoupled or simply absent there are only 
six complex Yukawa couplings and, at first sight, the formula 
(\ref{eqn:seesaw}) should determine all of them unambiguously. 
However, it is easy to see that the resulting set of equations
\beq
\label{eqn:equations}
\xi_A^2 \delta^{AB}+\sum_Cx_C^2 \YY{CA} \YY{CB}=0~,
\phantom{aaa}A,B=1,2,3
\eeq
with $C=1,2$ is self-consistent only if $m_{\nu_1}\propto\xi_1^2=0$. In other 
words, the absence or decoupling of one right-chiral neutrino automatically 
ensures det$(\mathbf{m}_\nu)=0$, that is $m_{\nu_1}=0$ (with only one 
right-chiral neutrino two eigenvalues of $\mathbf{m}_\nu$ would vanish) and, 
hence, hierarchical or inversely hierarchical spectrum of the left-chiral 
neutrinos. For $m_{\nu_1}=0$ not all equations (\ref{eqn:equations}) are 
independent; as a result there is still a two-parameter freedom which can be 
parametrized by $\YY{22}$. Since for one right-chiral neutrino (no summation
over $C$ in eq. (\ref{eqn:equations})) all Yukawa couplings are unambiguously
determined by the formula (\ref{eqn:seesaw}), it follows, that for 
three right-chiral neutrinos a convenient parametrization of the ambiguity 
is given by the $\YY{22}$, $\YY{32}$ and $\YY{33}$, as it allows to take the 
limit in which heavy right-chiral neutrinos are succesively decoupled. 
From the above it is clear that the description of leptogenesis simplifies 
enormously in the decoupling limit for which the first condition is the 
sufficiently small value of $m_{\nu_1}$ and, hence, the spectrum of 
left-chiral neutrinos close to hierarchical or inversely hierarchical.

For $x_3\YY{32}=0$ and $x_3\YY{33}=0$ the exact solution of the equations 
(\ref{eqn:equations}) in terms of the matrix $\mathbf{\Omega}$ reads
\beq
\label{eqn:Omega}
\mathbf{\Omega}=\left(\matrix{0&\pm i\sqrt{-1-z^2}& iz\cr
                              0&-iz&\pm i\sqrt{-1-z^2}\cr
                              1& 0 &        0            }\right)~,
\phantom{aaa} {\rm where}\phantom{aa} z=(x_2/\xi_2)\YY{22}~.
\eeq 
It is obvious that all non-vanishing quantities: $\YY{12}$, $\YY{13}$ and 
$x_2\YY{23}$ depend then only on $x_2\YY{22}$. The 
decoupling of the third right-chiral neutrino is a somewhat stronger 
requirement as it corresponds to $x_3\YY{3A}\to 0$ for all $A=1,2,3$. In 
agreement with our previous discussion this requires one massless 
left-chiral neutrino because, as follows from eqs. 
(\ref{eqn:Ygensol}, \ref{eqn:Omega})
\beq
\label{eqn:xi31}
x_2\YY{31} = i\xi_1~.
\eeq
While the results for the baryon asymmetry described below depend only 
on the particular form (\ref{eqn:Omega}) of the matrix $\mathbf{\Omega}$, 
it is clear that the most natural way to realize the limit $x_3\YY{32}=0$ 
and $x_3\YY{33}=0$ is to have $x_3\to 0$ and therefore also one nearly 
massless left-chiral neutrino.

\begin{figure}
\epsfig{figure=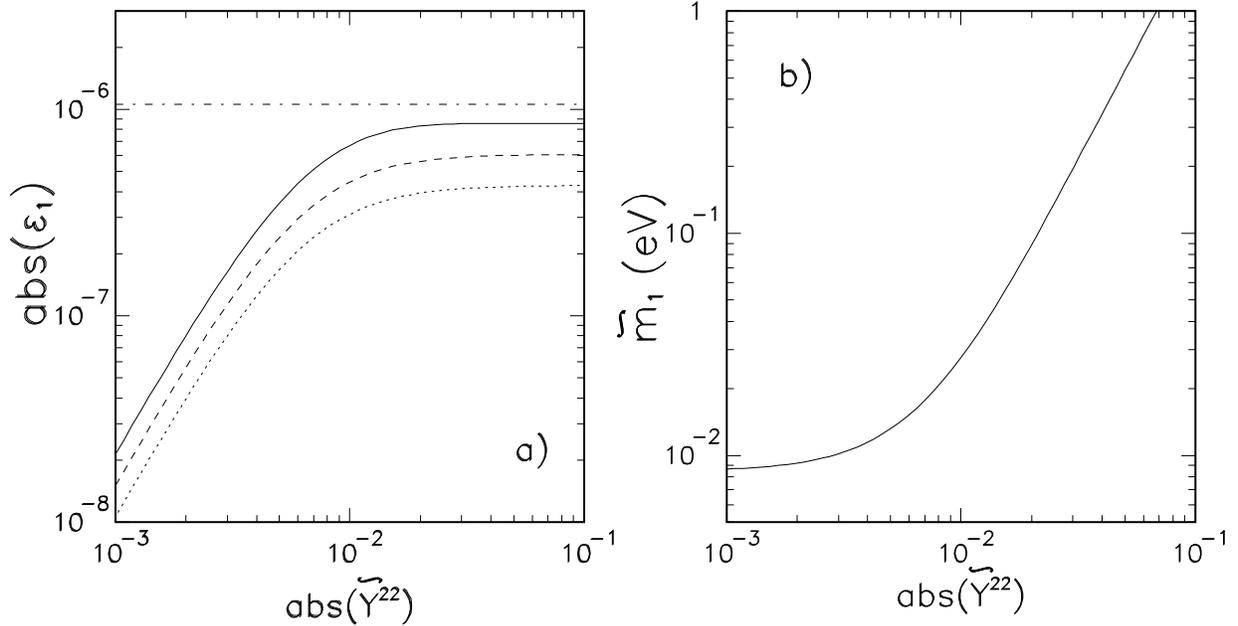,width=\linewidth} 
\vspace{1.0truecm}
\caption{\protect $|\epsilon_1|$ and $\tilde m_1$ as a function of 
$|\tilde\mathbf{Y}^{22}|$ for $M_1=10^{10}$ GeV, $x_2=0.1$
and strictly hierarchical light neutrino masses. Solid, dashed and
dotted lines correspond to $\varphi$=arg$(\tilde\mathbf{Y}^{22})=\pi/4$, 
$\pi/8$  and $\pi/12$, respectively. $\tilde m_1$ is insensitive to $\varphi$.
The dash-dotted line shows $|\epsilon_1|^{\rm max}$ from \cite{DAIB}.}
\label{fig:hlpg1}
\end{figure}

Consider now the scenario with $x_2\simlt0.3$, in which the decays of the 
second right-chiral neutrino do not contribute to the generated lepton 
asymmetry. In this limit only the asymmetry parameter $\epsilon_1$ is 
relevant and can be approximated by\footnote{This is for the 
non-supersymmetric case; in this limit supersymmetric $\epsilon_1$ is 
twice as big \cite{COROVI}.} \cite{DAIB}
\beq
\label{eqn:epsilon}
\epsilon_1\approx-{3\over16\pi}{2M_1\over v^2}
{\sum_A{\rm Im}\left(\mathbf{\Omega}^2_{1A}\right)m^2_{\nu_A}\over
\sum_A\left|\mathbf{\Omega}_{1A}\right|^2m_{\nu_A}}~.
\eeq
For $\mathbf{\Omega}$ given in (\ref{eqn:Omega}) it is easy to find that
\beq
\epsilon_1\approx\left\{\matrix{
-(3/16\pi) (\xi_3/\xi_2)^4 |x_2\YY{22}|^2 \sin(2\varphi) 
& {\rm for} & |z|\ll \xi_2/\xi_3 \cr
-(3/16\pi) (\xi_3^2-\xi_2^2) \sin(2\varphi) & {\rm for} & |z|\gg1}
\right.~,
\eeq
where $\varphi={\rm arg}(\YY{22})$. For hierarchical masses of the 
left-chiral neutrinos this is illustrated in fig.~\ref{fig:hlpg1}a where 
we plot $|\epsilon_1|$ as a function of $\YY{22}$ for fixed $M_1$ and $x_2$. 
The maximal value 
$|\varepsilon_1|^{\rm max}=(3/16\pi)(2M_1/v^2)(m_{\nu_3}-m_{\nu_2})$ 
attained in the decoupling limit is about 20\% lower than the upper limit 
$|\varepsilon_1|^{\rm max}=(3/16\pi)(2M_1/v^2)\sqrt{\Delta m_\mathrm{atm}^2}$ 
in the general hierarchical case derived in \cite{DAIB} and used in 
\cite{BUDIPL}. In the inversely hierarchical case, which is also compatible 
with the decoupling limit, $|\epsilon_1|$ is generically smaller, as then 
$m_{\nu_3}-m_{\nu_2}\simlt 
(\Delta m_{\rm sol}^2/2\sqrt{\Delta m_{\rm atm}^2})\ll 
\sqrt{\Delta m_{\rm atm}^2}$.

\begin{figure}
\epsfig{figure=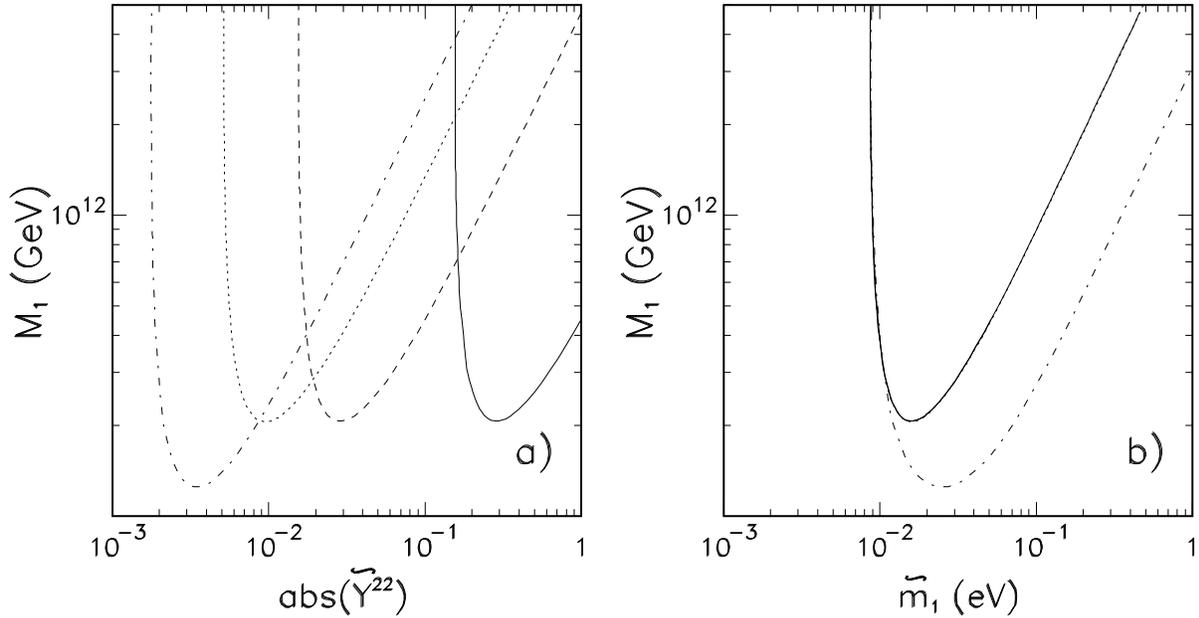,width=\linewidth} 
\vspace{1.0truecm}
\caption{Lower limits on $M_1$ for $x_2=0.01$,
$0.1$, $0.3$ and $0.95$ (solid, dashed, dotted and dash-dotted lines,
respectively) as functions of $|\tilde\mathbf{Y}_\nu^{22}|$ (panel a) 
and $\tilde m_1$ (panel b).}
\label{fig:hlpg2}
\end{figure}

Efficient generation of the lepton number asymmetry requires not only 
a sufficiently large value of $|\epsilon_1|$, but also that the lightest 
right-chiral neutrinos decay out of equilibrium. This requirement can 
be expressed as the condition that their decay rate $\Gamma_1$ must be 
smaller than the Hubble parameter $H$ at the temperature $T\sim M_1$ 
\cite{GIFIPA,PLUMIS}. This in turn translates into the following 
condition \cite{DAIB}:
\beq
\label{eqn:outofeqcond}
\tim\equiv\sum_A |\OD{1A}| m_{\nu_A} \simlt 5\times 10^{-3} ~{\rm eV}~.
\eeq
For $\tim$ of this order and larger the wash-out processes are very 
efficient and suppress the resulting lepton asymmetry \cite{PLUMIS,BUDIPL}. 
In the decoupling limit the condition (\ref{eqn:outofeqcond}) can hardly be 
satisfied, because, as follows from the form (\ref{eqn:Omega}) of the matrix 
$\mathbf{\Omega}$, one has $\tim >m_{\nu_2}\simgt8\times 10^{-3}$~eV in 
the hierarchical case (for inverse hierarchy $\tim\simgt0.05$~eV). Moreover, 
the smallest possible values of $\tim$ are obtained for $|z|\ll 1$ where 
$\epsilon_1$ is suppressed (see fig.~\ref{fig:hlpg1}). As a result, in order 
to reproduce the observed baryon asymmetry through the leptogenesis the mass 
$M_1$ of the lightest right-chiral neutrino has to be about two orders of
magnitude bigger than its lower limit found in \cite{BUDIPL} for general 
see-saw models leading to the hierarchical spectrum of light neutrinos. This 
is illustrated in fig.~\ref{fig:hlpg2}  where the lower limits on $M_1$ 
(resulting from the requirement that the computed $B-L$ abundance $Y_{B-L}$ 
is not smaller than $2.8\times10^{-10}$) are plotted for $x_2=0.01$ (solid 
line), $x_2=0.1$ (dashed line), $x_2=0.3$ (dotted line) and $x_2=0.95$ 
(dash-dotted line) as functions of $|\YY{22}|$ (panel a) and $\tim$ (panel b).
In agreement with our discussion, for $x\simlt 0.3$, when the lepton number 
asymmetry arises entirely from the lightest neutrino decays, the limits on 
$M_1$ depend only on $x_2\YY{22}$ that is on $\tim$ (the three curves 
corresponding to $x_2=0.01$, $0.1$ and $0.3$ merge into a single one when 
plotted against $\tim$) and not on $x_2$ and $\YY{22}$ separately. It is 
worth stressing that in the range of $\tmi$ values accessible in the 
considered limit, the prediction for the baryon asymmetry of the Universe 
is almost insensitive to the initial abundance of the first right-chiral 
neutrinos \cite{PLUMIS,BUDIPL}. One should also note that our results are
consistent and generalize those obtained in ref. \cite{RAST} in the specific
see-saw model with exactly two right-chiral neutrinos.

\begin{figure}
\epsfig{figure=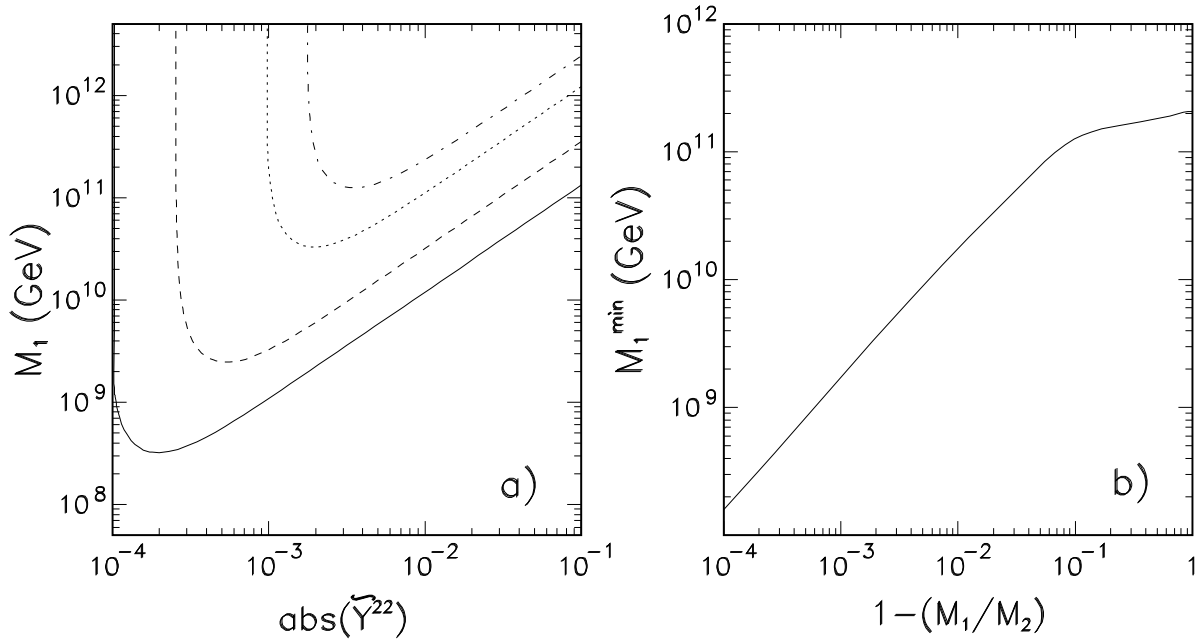,width=\linewidth} 
\vspace{1.0truecm}
\caption{Lower limits on $M_1$. Panel a):
as functions of $|\tilde\mathbf{Y}_\nu^{22}|$ 
for $x_2=0.95$, $0.99$, $0.999$ and $0.9999$ (dash-dotted, dotted, dashed, 
and solid lines, respectively). Panel b): as a function of the degeneracy
of the two right-chiral neutrinos.}
\label{fig:hlpg3}
\end{figure}

A non-trivial dependence on $x_2$ and $\YY{22}$ separately can arise 
only through direct contribution of the second heavy neutrino to the 
generated lepton asymmetry. This is shown if fig.~\ref{fig:hlpg3}a where 
we plot the lower limit on $M_1$ for the case of degenerate two right-chiral 
neutrinos.\footnote{For $|\YY{22}|\sim0.1$ the curve corresponding to 
$x=0.9999$ underestimates slightly the lower limit on $M_1$ because there
the applicability condition $\Gamma_1\ll M_2-M_1$ for the formulae for 
$\epsilon_{1,2}$ is not satisfied and one should use more general
expressions derived in \cite{PILUS}.}
The dependence of this lower limit on the degeneracy of the two right-chiral 
neutrinos is shown in fig.~\ref{fig:hlpg3}b. In agreement with the 
observation made in ref. \cite{ELRAYA1}, the degeneracy of two neutrinos 
allows to significantly lower the reheat temperature. Linear dependence of 
$M_1^{\rm min}$ on the degeneracy parameter $\delta=1-M_1/M_2=1-x^2_2$ seen 
in fig.~\ref{fig:hlpg3}b for $\delta<0.05$ is related to the fact that for
degenerate two right-chiral neutrinos both, $\epsilon_1$ and $\epsilon_2$,
are proportional to $M_1/\delta$ \cite{BRetAL}. From fig.~\ref{fig:hlpg3} it 
follows that in this framework lowering $M_1$ below $10^9-10^8$ GeV can be 
achieved if $\delta\simlt10^{-3}-10^{-4}$. The stability of such 
a degenerate mass pattern is however an open question. Another possibility 
is to consider a non-thermal production of right-chiral neutrinos and $-$ in 
the supersymmetric scenario $-$ sneutrinos \cite{GIPERITK}. A considerably 
lower reheat temperature is also sufficient in the very interesting scenario 
proposed in \cite{MUYA} and elaborated recently in \cite{ELRAYA2},
in which the $B-L$ asymmetry is produced directly by the decays of 
right-chiral sneutrinos playing the role of the inflaton.

The results described above apply for $x_3\YY{32}\to0$, $x_3\YY{33}\to0$.
Solving explicitly the set of equations (\ref{eqn:equations}) with $C=1,2,3$
in terms of $\YY{22}$, $\YY{32}$ and $\YY{33}$ it is easy to see (we 
have also checked it numerically) that as long as
\beq
\label{eqn:deccond}
x_3|\YY{3A}|<{\cal O}(0.1)~{\rm min}\left\{x_2|\YY{22}|,~\xi_A\right\}~,
\phantom{aaa}{\rm for}\phantom{aa}A=2,3	
\eeq
the resulting values of $\epsilon_1$ and $\tmi$ and, hence, the
predictions for the $B-L$ abundance are almost unchanged. In concrete 
theoretical models of the see-saw mechanism these conditions can be 
satisfied either if the independent couplings $\YY{32}$ and $\YY{33}$ are 
sufficiently weak (because $\YY{3A}=\sum_B \mathbf{Y}_\nu^{3B}\mathbf{U}^{BA}$
and taking into account the bi-large structure of the matrix $\mathbf{U}$, 
this condition is automatically satisfied if the couplings 
$\mathbf{Y}_\nu^{3A}$ are small) or, more naturally, for the third 
right-chiral neutrino sufficiently heavy (typically in such models 
$M_3\simgt10^{16}$~GeV). 

In the regime $x_2\simlt0.3$ when only decays of the first right-chiral 
neutrino contribute to the $B-L$ asymmetry, the observed baryon asymmetry 
can be generated for $M_1$ as low as $\simlt10^9$~GeV \cite{BUDIPL} only if 
the conditions (\ref{eqn:deccond}) are strongly violated. To see this 
explicitly, we note that this is possible only if $\tim$ can be made 
substantially smaller than $m_{\nu_2}$ \cite{BUDIPL} keeping at the same 
time not too small $|\epsilon_1|$. From the formulae (\ref{eqn:epsilon}), 
(\ref{eqn:outofeqcond}) and the orthogonality relation 
$\sum_B\mathbf{\Omega}_{1B}^2=1$ it follows that this requires 
$\mathbf{\Omega}_{11}^2\approx1$ and, for $A=2,3$, 
$\mathbf{\Omega}_{1A}^2\sim{\cal O}(m_{\nu_1}/m_{\nu_A})$ with nonnegligible
imaginary part. The orthogonality relation applied to the first column 
of $\mathbf{\Omega}$ then implies that $\mathbf{\Omega}_{31}^2$ is at most 
of order ${\cal O}(m_{\nu_1}/m_{\nu_3})$. Since (using again the 
orthogonality relation)
\begin{eqnarray}
\mathbf{\Omega}_{31}^2=1-\mathbf{\Omega}_{32}^2
-\mathbf{\Omega}_{33}^2
=1-(x_3/\xi_2)^2(\tilde\mathbf{Y}^{32}_\nu)^2
-(x_3/\xi_3)^2(\tilde\mathbf{Y}^{33}_\nu)^2~,
\end{eqnarray}
such a suppression of the element $\mathbf{\Omega}_{31}$  requires
$\tilde\mathbf{Y}^{32}_\nu$ and/or $\tilde\mathbf{Y}^{33}_\nu$ violating the 
conditions (\ref{eqn:deccond}). At least one of the couplings 
$(\tilde\mathbf{Y}^{3A}_\nu)^2$ must be then of order
\begin{eqnarray}
(\tilde\mathbf{Y}^{3A}_\nu)^2\sim3.3\left({m_{\nu_A}\over{\rm eV}}\right)
\left({M_3\over10^{14}~{\rm GeV}}\right)\phantom{aaa}A=2\phantom{a}{\rm or}
\phantom{a}A=3~,
\end{eqnarray}
that is, it approaches the nonperturbative regime for $M_3\simgt10^{16}$~GeV.

It is also interesting to note that in the decoupling limit the requirement 
of perturbativity of the neutrino Yukawa couplings imposes some additional 
constraints on the neutrino masses. From the estimate 
$\xi_A^2=3.3\times10^{-4}(m_{\nu_A}/{\rm eV})(M_1/10^{10}~{\rm GeV})$ and 
(\ref{eqn:xi31}) one gets:
\beq
\left(\YY{31}\right)^2 \approx3.3\left({m_{\nu_1}\over{\rm meV}}\right)
\left({M_3\over10^{17}~{\rm GeV}}\right)~.
\eeq
Therefore, for $m_{\nu_1}\approx10^{-3}$~eV the mass of the heaviest 
right-chiral neutrino cannot exceed $10^{17}$~GeV. Similarly, for 
$M_2\gg M_1$, that is for $x_2\ll 1$, when $|\OO{23}|\approx 1$, the 
requirement of perturbativity of $\YY{23}$ leads to the bound 
$M_2\simlt 10^{15}$~GeV.

In conclusion, if one of the right-chiral neutrinos 
effectively decouples from the see-saw mechanism the resulting spectrum 
of the left-chiral light neutrinos must be hierarchical (or inversely 
hierarchical). Then, if the baryon number asymmetry of the Universe is to 
be generated by the decays of the lightest right-chiral neutrino only, 
its mass has to be higher than $\sim2\times10^{11}$~GeV irrespectively
of its initial abundance. Therefore, if the decaying neutrinos are produced 
thermally the required reheat temperature 
$T_{\rm reh}$ has to be $\simgt10^{12}$~GeV. The decoupling limit is 
practically realized in a broad class of theoretical see-saw models invoked 
to reconcile the hierarchical spectrum of light neutrinos with their 
bi-large mixing pattern. Supersymmetric versions of such models face 
therefore the serious problem of gravitino overproduction if the 
right-chiral neutrinos are produced thermally.

\vskip0.3cm
\noindent{\bf Acknowledgments} We would like to thank S. Pokorski for 
many fruitful discussions and reading the manuscript and M. Raidal for 
comments and bringing ref. \cite{KIRO} to our attention. The work was 
partially supported by the EC contract HPRN-CT-2000-0148 and the Polish 
State Committee for Scientific Research grants 2 P03B 040 24 for 
2003-2005 (P.H.Ch.) and 2 P03B 129 24 for 2003-2005 (K.T.).


\begin{thebibliography}{99}

\bibitem{WMAP} D.N. Spergel et al. (the WMAP Collaboration), 
               astro-ph/0302209.

\bibitem{GERASL} M. Gell-Mann, P. Ramond and R. Slansky, in {\sl Proceedings 
                 of the Supergravity Stony Brook Workshop}, New York, 1979 
                 (eds. P. van Nieuvenhuizen and D.Z. Freedman, North-Holland, 
                 Amsterdam); T. Yanagida, in {\sl Proceedings of the Workshop 
                 on Unified Theories and Baryon Number in the Universe}, 
                 Tsukuba, Japan, 1979 (eds. A. Sawada and A. Sugamoto, 
                 KEK Report No. 79-18, Tsukuba); R.N. Mohapatra and G.
                 Senjanovi\'c, {\sl Phys. Rev. Lett.} {\bf 44} (1980) 912. 

\bibitem{CHPO} P.H. Chankowski and S. Pokorski, {\sl Int. J. Mod. Phys.}
               {\bf A17} (2002) 575 and references therein.

\bibitem{FUYA} M. Fukugita and T. Yanagida, {\sl Phys. Lett.} {\bf B174}
               (1986) 45.

\bibitem{ALFE} G. Altarelli and F. Feruglio, in {\sl Neutrino Mass}
               Springer Tracts in Modern Physics, G. Altarelli and
               K. Winter eds. (hep-ph/0206077);
               M.C. Gonzales-Garcia and Y. Nir, hep-ph/0202058.

\bibitem{ALBA} C.H. Albright and S.M. Barr, {\sl Phys. Rev.} {\bf D58}
               (1998) 013002;
               C.H Albright, K.S. Babu and S.M. Barr, {\sl Phys. Rev. Lett.}
               {\bf 81} (1998) 1167;
               P.H. Frampton and A. Rasin, {\sl Phys. Lett.} {\bf B478}
               (2000) 424;
               N. Irges, S. Lavignac and P. Ramond, {\sl Phys. Rev.} {\bf D58}
               (1998) 035003;
               G. Altarelli and F. Feruglio, {\sl Phys. Lett.} {\bf B439}
               (1998) 112, {\sl JHEP} {\bf 9811} (1998) 021.

\bibitem{KI1} S.F. King, {\sl Phys. Lett.} {\bf B439} (1998) 350, 
              {\sl Nucl. Phys.} {\bf B562} (1999) 57;
              S. Davidson and S.F. King, {\sl Phys. Lett.} {\bf B445}
              (1998) 191;
              Q. Shafi and Z. Tavartkiladze, {\sl Phys. Lett.} {\bf B451}
              (1999) 129.

\bibitem{KI2} S.F. King, {\sl Nucl. Phys.} {\bf B576} (2000) 85.

\bibitem{LAMASA} S. Lavignac, I. Masina and C.A. Savoy, {\sl Phys. Lett.} 
                 {\bf B520} (2001) 269, {\sl Nucl. Phys.} {\bf B633} 
                 (2002) 139.

\bibitem{KIRO} S.F. King and G.G. Ross, {\sl Phys. Lett.} {\bf B520}
                (2001) 243.

\bibitem{RAST} A. Strumia and M. Raidal, {\sl Phys. Lett.} 
               {\bf 553} (2003) 72. 

\bibitem{CAIB} J.A. Casas and A. Ibarra, {\sl Nucl. Phys.} {\bf B618}
               (2001) 171.

\bibitem{COROVI} L. Covi, E. Roulet and F. Vissani, {\sl Phys. Lett.} 
                 {\bf B384} (1996) 169.

\bibitem{DAIB} S. Davidson and A. Ibarra, {\sl Phys. Lett.} {\bf B535}
               (2002) 25.

\bibitem{BUDIPL} W. Buchm\"uller, P. Di Bari and M. Pl\"umacher, 
                 {\sl Nucl. Phys.} {\bf B643} (2002) 367,
                 {\sl Phys. Lett.} {\bf B547} (2002) 128,
                 preprint CERN-TH/2003-016 (hep-ph/0302092).

\bibitem{BRetAL} G.C. Branco {\em et al.}, {\sl Phys. Rev.} {\bf D67} 
                 (2003) 073025.

\bibitem{PLUMIS} M. Pl\"umacher, {\sl Z. Phys.} {\bf C74} (1997) 549.

\bibitem{PILUS} A. Pilaftsis, {\sl Int. J. Mod. Phys.} {\bf A14} (1999) 1811.

\bibitem{ELRAYA1} J. Ellis, M. Raidal and T. Yanagida, {\sl Phys. Lett.} 
                 {\bf B546} (2002).

\bibitem{GIFIPA} G.-F. Giudice, W. Fishler, R.G. Leigh and S. Paban,
                 {\sl Phys. Lett.} {\bf B258} (1991) 45.

\bibitem{GIPERITK} G.-F. Giudice, M. Peloso, A. Riotto and I. Tkachev, 
                   {\sl JHEP} {\bf 9908} (1999) 014.

\bibitem{MUYA} H. Murayama and T. Yanagida, {\sl Phys. Lett.} {\bf B322} 
               (1994) 349; K. Hamaguchi, H. Murayama and T. Yanagida, 
               {\sl Phys. Rev.} {\bf D65} (2002) 043512; T. Moroi and 
               H. Murayama, {\sl Phys. Lett.} {\bf 553} (2003) 126.

\bibitem{ELRAYA2} J. Ellis, M. Raidal and T. Yanagida, preprint 
                  CERN-TH/2003-073 (hep-ph/0303242).

\end{thebibliography}
\end{document}